\documentclass[twocolumn,showpacs,preprintnumbers,amsmath,amssymb,pre]{revtex4}
\usepackage{amssymb,amsmath,amscd}
\usepackage{epsfig}
\usepackage{bm}
\def\beqa{\begin{eqnarray}}
\def\eeqa{\end{eqnarray}}
\def\a={&=&}
\newcommand{\half}{\frac{1}{2}}   
\newcommand{\pa}[1]{\left( {#1} \right)}
\newcommand{\pas}[1]{\left[ {#1} \right]}
\newcommand{\ave}[1]{\langle {#1} \rangle}
\def\nnn{\nonumber \\}
\def\cia{c_{i-1}} 
\def\cib{c_{i+1}} 
\def\ci{c_i} 
\begin{document}
\title{
Emergence of Cooperation with Self-organized Criticality
}
\author{Hyeong-Chai Jeong and Sangmin Park}
\affiliation{
Department of Physics, Sejong University, Seoul 143-747, Korea, 
}
\date{\today}
 
\begin{abstract}
Cooperation and self-organized criticality are two main keywords in 
current studies of evolution. We propose a generalized Bak-Sneppen model
and provide a natural mechanism which accounts for both phenomena 
simultaneously. We use the prisoner's dilemma games to mimic 
the interactions among the members in the population.
Each member is identified by its cooperation probability, 
and its fitness is given by the payoffs from neighbors. 
The least fit member with the minimum payoff is replaced by 
a new member with a random cooperation probability. 
When the neighbors of the least fit one are also replaced with 
a non-zero probability, a strong cooperation emerges.
The Bak-Sneppen process builds a self-organized structure
so that the cooperation can emerge even in the parameter 
region where a uniform or random population decreases the number of
cooperators.  
The emergence of cooperation is due to the same dynamical 
correlation that leads to self-organized criticality in 
replacement activities.
\end{abstract} 


\maketitle

\section{Introduction}
A fundamental question in the theory of evolution has been
how cooperation can emerge between selfish
members~\cite{Nowak06B,Nowak:sci.06,Milinski87,Segbroeck09,Gomez07,Parcheco06,Santos05}.  
Another question is why evolution takes place in terms of
intermittent bursts of activities, which are the characteristics of 
dynamical systems in a `critical' state~\cite{Gould77,Raup86}. 
Here, we propose a generalized Bak-Sneppen (BS) model~\cite{Bak93},
which may solve the above two puzzles simultaneously.
We take an approach of evolutionary game theory and use 
the prisoner's dilemma (PD) games to mimic 
the interactions among members. 
Each member is identified by its stochastic strategy,
specified by its (history independent) cooperation 
probability (CP). 
Here, a `member' can represent an individual in a species, an agent in
an economical system or a species in an ecological system.   
The fitness of a member is given by 
the payoffs of the games with its neighbors.   
We then apply BS dynamics and replace 
the least fit member and its neighbors by new
members with random CPs.
The neighbors of the non-cooperator are likely to vanish
due to its low payoff, but the non-cooperator itself can 
also be removed through the BS mechanism. 
As the non-cooperators disappear, the overall CP increases,
and a new comer (with a random CP) will have a lower CP than 
the increased average. 
Therefore, the new comer tends to cause its neighbor to be the
least fit, and the replacement activity likely occurs at or near
the new comer's site. This invokes the spatio-temporal correlation 
between the least fit sites and can explain 
why replacements are episodic as well as how cooperation emerges.

Evolutionary game theory has been one of the most powerful tools in
studying the dynamics of evolution~\cite{Nowak06B}. 
However, a simple straightforward application of game theory
cannot explain the strong cooperation between ``selfish" 
replicators observed in nature and society.   
For the evolution to construct a new, upper level of
organization, cooperation amongst the majority of the population 
is needed. However, the game theoretical description of 
interactions between members usually leads to defections as evolutionarily stable
strategies. Natural selection, which has been a fundamental 
principle of evolution, prefers the species that
beat off the others and oppose cooperation.

There have been numerous studies looking for natural mechanisms for
the evolution of cooperation among competitive 
members~\cite{Hamilton64,Milinski87,nowak:nat.98,West02}.   
Recently, Nowak presented a state-of-art review on the evolution of
cooperation and discussed five known mechanisms: 
kin selection, direct reciprocity, indirect reciprocity,   
network reciprocity, and group selection~\cite{Nowak:sci.06,Grafen:JEB.07}.  
Extensive studies provide the exact conditions for the emergence of
cooperation for each of the five mechanisms. However, such conditions 
do not seem to be general enough to explain 
the cooperative phenomena observed everywhere.
For example, for network reciprocity, the benefit-to-cost 
ratio of a cooperative behavior should be larger than the average 
degree~\cite{Nowak:sci.06}, but this seems to be a rather strong assumption  
because the degrees are quite large in most cases in real population
structures. 
Also, there have been a great deal of studies on self-organized criticality in game 
theory~\cite{Scheinkman:AMR94,Sole:JTB95,Killingback:JTB98,Arenas:EDC02,Ebel:PRE02},
but their dynamics leading to the critical states are not directly
connected to the emergence of cooperation. 
Here, we consider an evolutionary game on networks
and show that cooperation can emerge when the benefit to cost ratio
is larger than just 1 if we use the BS process. 
When cooperators interact with defectors, 
they tend to disappear, giving rise to an assortment of
cooperators~\cite{Fletcher:PRSB.09}. Furthermore, this behavior emerges
in the long run even with a small ``chain-death'' rate, $\omega$, where the 
number of neighbors that get replaced is less than one. For a uniform or
random arrangement of cooperators and defectors, more cooperators than
defectors disappear for small $\omega$, but in the long run, the BS process
builds a self-organized structure so that the number of cooperators 
in the population increases. 

\section{model}
An influential model aimed to mimic the interactions 
between competitive members in a population 
is the PD game. 
It is one of the matrix games between two players who have 
two possible decisions, cooperation ($C$) or
defection ($D$). 
We consider a case in which the payoffs are calculated by 
the cost $c$ and the benefit $b$ of a cooperative behavior.
If one player defects while the other cooperates, 
the defector receives benefit $b$ without any cost
whereas the cooperator pay cost $c$ and its payoff becomes $-c$.
For mutual cooperation, both get benefit $b$, but pay cost $c$, and
their payoffs become $b-c$ while the payoffs for 
mutual defection are 0. 
When we add $c$ to all elements so that payoff can be 
directly interpreted as (non-negative) fitness,
the payoff matrix becomes
\[
\begin{array}{cc}
  & \ \ \begin{array}{cc}
     C & D  
      \end{array} \ \	\\
 \begin{array}{c}
      C      \\
      D   
 \end{array}
 & \left( \begin{array}{cc}
      b     &   0   \\
      b+1   &   1   \\
  \end{array} \right),  \\
\end{array}
\] 
where we set $c=1$ without loss of generality.
With conventional competition processes,
the matrix game shown above
does not, in general, predict the evolution of
cooperation.
The birth-death process
always predicts an evolution of defection. 
Cooperation can emerge for death-birth or imitation processes in
a structured population, but only with a (unrealistic) large value of
the benefit-to-cost ratio $b$ for real populations~\cite{Nowak:sci.06}. 

Here, we consider the PD game interaction,
but introduce the BS mechanism~\cite{Bak93}
as the competition process,
and assume that the least fit member and
its neighbors are prone to disappear.
Each member is characterized by its strategy that determines
when to choose the `decisions' $C$ or $D$. 
We consider the history-independent stochastic strategies, 
and the phenotype of a member, say the $i$th member,
is represented by its CP $c_i$. 
The history independent pure (deterministic) strategies, 
the ``always $C$'' and  the ``always $D$'', correspond to the 
limits of $c_i=1$ and $c_i=0$, respectively. 
The fitness of a member is  given by the sum of payoffs
from its neighbors, and the member dies out
if its total payoff is the minimum. The died-out site
is occupied by a new member with a new CP,
which is drawn randomly from 0 to 1. 
Neighbors of the least fit site may also be harmed 
in the process of establishing the steady interaction 
with the new comer. Hence, we replace the
neighbors of the least fit site by new members with the ``chain-death'' 
probability $\omega >0$.

%%%%%%%%%%%%%%%%%%%%%%%%(Fig 1: OSDS)%%%%%%%%%%%%%%%%%%%%%%
\begin{figure}[t!] 
\includegraphics[width=7.0cm]{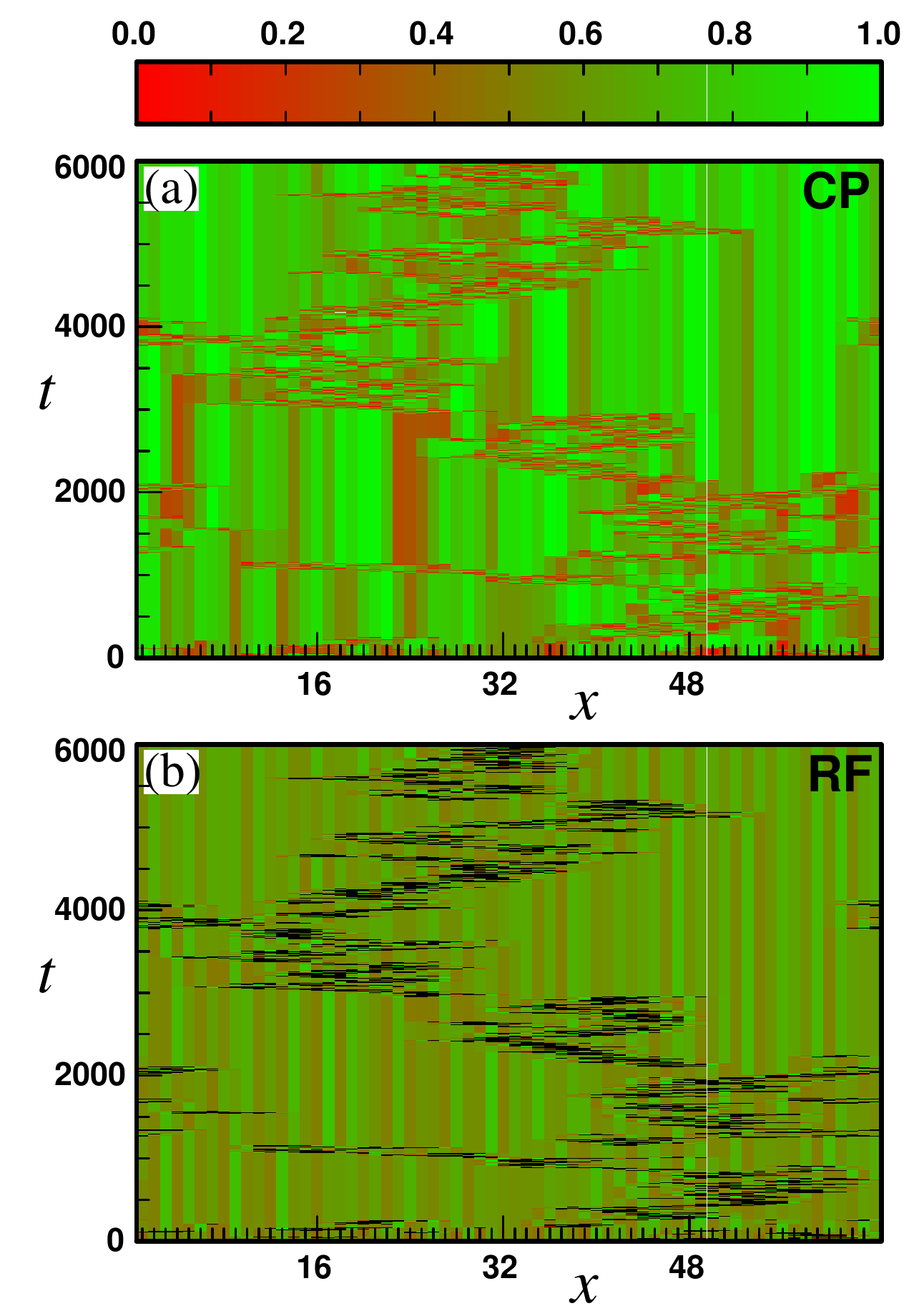} 
\caption[0]{ Real space configurations of
(a) the CP $c_i$ and (b) the RF $\tilde{f}_i$ 
for $t \in [0, 6000]$ with $\omega=1$ and $N=64$.
They are represented by colors, red for~0 and green for~1,
as indicated by the top panel. 
The black dots in (b) represent the least fit sites. 
}
\label{f.1}
\end{figure}
%%%%%%%%%%%%%%%%%%%%%%%%%%%%%%%%%%%%%%%%%%%%%%%%%%%%%%%%%%%%

\section{Methods and Results}
We study the strategy evolution of a simple structured population
from the initial state of random strategies.  
Initially, members in the population have cooperation probabilities
that are drawn randomly from the uniform distribution of the interval
[0\ 1]. They play PD games with their nearest neighbors. 
We assume that each member plays sufficiently many games prior to the
reproduction process and use the payoff expectation value as its
fitness. The least fit member with the minimum payoff expectation is
replaced by a new member with a new random CP. 
In addition to the least fit member, the neighbors of the least fit
member are also replaced by new members with the
probability~$\omega$. 
Then, we recalculate the payoff expectations,
and replacements occur at the new least fit member and its neighbors.
We continue these processes until the system reaches a steady state
and calculate the statistical properties of the population, such as 
the mean cooperation probability, fitness distribution, 
avalanche size (defined later) distribution and etc. 

For simplicity, we present our model and results in a one-dimensional (1D)
structure, but our main results hold in other population structures. 
Initially ($t=0$), we assign a random CP, $c_i(0)$, 
to the site $i$ for $i=1,\ldots,N$. 
Then, we calculate the payoff expectation, $f_i(0)$, at time $t=0$, 
\beqa
 f_i(0) \a= b\,\pas{\cia(0)+\cib(0)} + 2 \pas{1-\ci(0)},
\label{e.fi}
\eeqa
of site $i$ with a periodic boundary condition and find the minimum
payoff site, $m_0$. Except this minimum site, $m_0$ and its neighbors,
$m_0\pm1$, the CPs are not changed at $t=1$, 
so we set $c_i(1)=c_i(0)$ unless $i=m_0$ or $m_0\pm1$. 
The CP at the $m_0$ site,
$c_{m_0}(1)$, is given by a new random number between 0 and 1. 
For its neighbor sites, 
$c_{m_0\pm1}(1)$ is given by a new independent random number
with the probability $\omega$, 
but remains as $c_{m_0\pm1}(0)$ with the probability $1-\omega$. 
Now, we recalculate the payoff $f_i$ of Eq.~(\ref{e.fi}) 
with $c_k(1)$ instead of $c_k(0)$.
We find the new minimum payoff site, $m_1$, of $t=1$ 
and apply the same replacement dynamics to get
$t=2$ configurations and so on.  

Figure~1 shows typical real space configurations 
of CP, $c_i$, and the reduced fitness (RF), 
$\tilde{f}_i = f_i/(2b+2) \in [0\ 1]$.
We show the configurations for initial  
6000 time steps of a $N=64$ system
with  $b=1.5$ and $\omega=1$.
Both the CP and the RF are represented by colors,
0 by red and 1 by green.
The least fit sites (black sites in~(b)) and 
their two neighbors are where the replacement activity occurs. 
Comparison between the configurations in~(a) 
and their equivalents in~(b) reveals that the least
fit sites are located where their neighbors are less
cooperative [relatively red in~(a)]. 
The disappearance of the ``red'' neighbors beside the 
least fit site by the BS-mechanism
shifts the overall system to green (more cooperative) with
time. 

%%%%%%%%%%%%%%%%%%% Use it later%%%%%%%%%%%%%%
%Once the system reaches to its steady state, 
%it shows avalanches of the least fit sites as 
%one can see from the correlation of the 
%black sites in~(b). 
%Green areas in Fig.~\ref{f.MF} increase with time
%indicating the evolution of cooperation. 
%%%%%%%%%%%%%%%%%%%%%%%%%%%%%%%%%%%%%%%%%%%%%%%

%%%%%%%%%%%%%%%%%%%%%%%%(Fig 2: OSDS)%%%%%%%%%%%%%%%%%%%%%%
\begin{figure}[t!] 
\includegraphics[width=8cm]{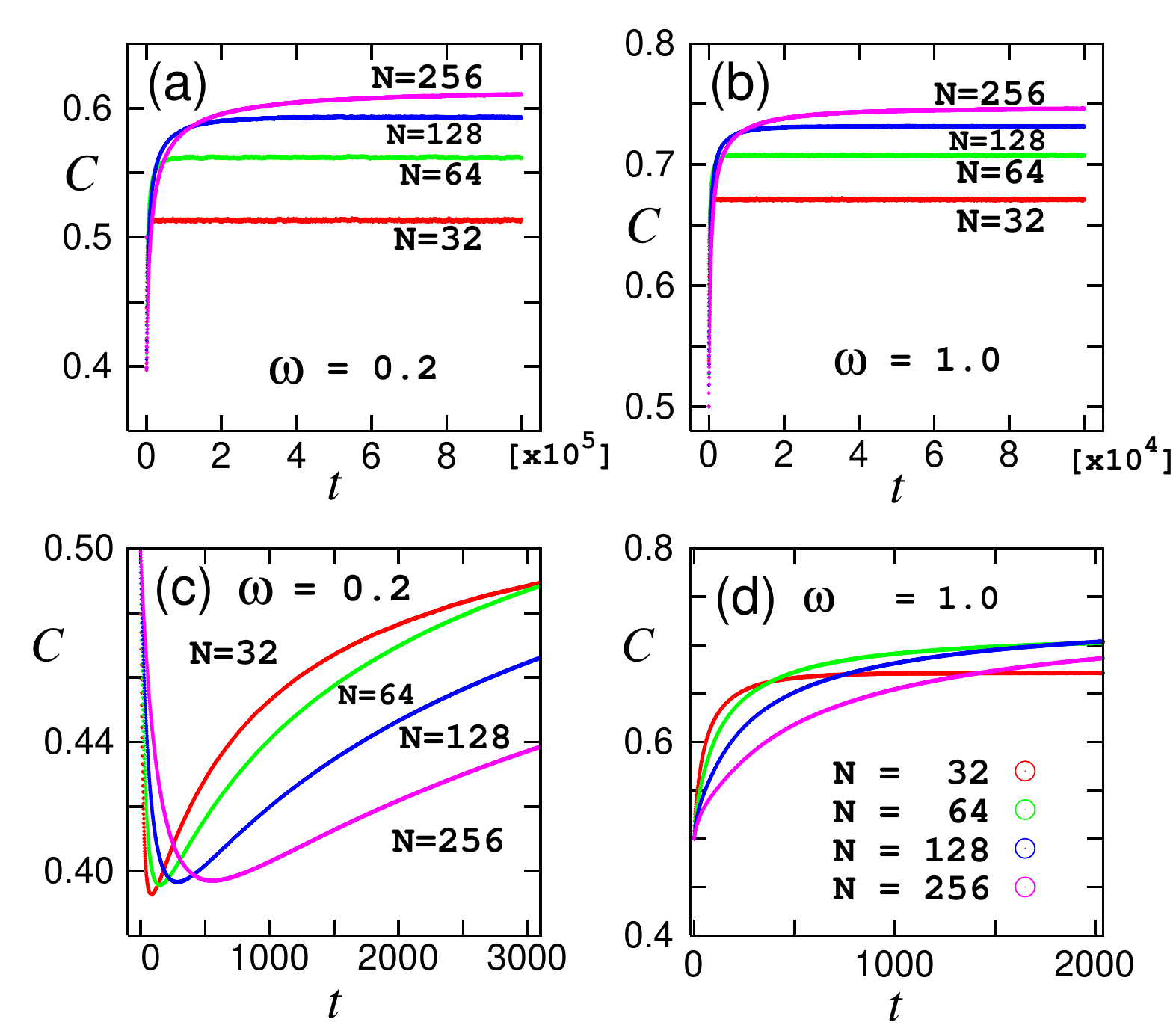} 
\caption[0]{
Time dependence of MCP, $C$, for (a,c) $\omega=0.2$ 
and (b,d) $\omega=1.0$ 
for systems with $N=32$, 64, 128, and 256.
In (a) and (b), the overall behaviors of MCP are shown while
the initial transient characteristics are shown in (c) and (d).
For $\omega=0.2$, MCP decreases first and then increases while it
monotonically increases for $\omega=1$. 
}
\label{f.2}
\end{figure}
%%%%%%%%%%%%%%%%%%%%%%%%%%%%%%%%%%%%%%%%%%%%%%%%%%%%%%%%%%%%

For a quantitative analysis, we measure the mean CP (MCP), 
$C(t) = \ave{\frac{1}{N} \sum_i c_i(t)}$, of the 
populations and show the results in Fig.~2.
Here, $\ave{\cdot}$ represents the ensemble average over many
different realizations of random initial configurations. 
Note that the MCP also represents the 
overall fitness $F(t) = \ave{\frac{1}{N} \sum_i f_i(t)}$ 
of the population because it is linearly related to MCP:
\beqa
 F(t) \a= \frac{1}{N}\ave{\sum_i  b\,\pas{\cia(t)+\cib(t)} 
           + 2  \pas{1-\ci(t)}}\nnn
      \a= 2 + 2 (b-1) C(t).
\eeqa
In Fig.~\ref{f.2}, the MCPs for four different system sizes, 
$N=32$, 64, 126, and 256, are shown
for two different values of $\omega$, $0.2$, and 1. 
We use $b=1.5$ for all figures in this paper, and all data are
obtained from numerical simulations.
Because we have assigned a random CP initially, 
the MCP starts from 0.5 at $t=0$. 
For $\omega=0.2$, the MCP decreases at
the beginning and then increases to the steady values
while it monotonically increases from the beginning for 
$\omega=1$, as shown in Figs.~2(c) and~(d).
Note that we have two different
elements in MCP changes. Replacement of 
the least fit member (which is likely to have a high CP)
tends to cause the MCP to decrease while the replacement of
its neighbors (which probably have low CPs) 
likely results in an increased MCP. 
The competition between these two elements governs the early dynamics
of the MCP.
It can decrease initially when 
$\omega < \omega_c \approx \frac{1}{k}  = 1/2$,
where $k$ is the number of neighbors. 
For a sufficiently large system, 
there would be a site, $m$, whose CP, $c_{m}$, is
arbitrarily close to one while those of its 
neighbors, $c_{m\pm 1}$, are almost zero. 
Hence, the expectation of MCP changes, 
$\Delta MCP$ would be 
$\frac{1}{N}\pas{\pa{\half-1}+k\omega\pa{\half-0}}$
and becomes negative for $\omega < \frac{1}{k}$ 
at the beginning. 
However, as time proceeds, the CP values develop
spatio-temporal correlations, and 
they govern the long-time dynamics.
Initially, the isolated high-CP cooperators 
are likely to be the least fit member,
and they are removed as time proceeds. 
Then, surviving cooperators 
remain in the groups,
and, thus, have high fitness. 
Now, low-CP defectors can be the least fit member,
especially when they are next to a very low-CP member.
The replacement of these low-CP member by new members with random CPs
causes the MCP to increase. Therefore, at a later time,
the MCP easily becomes larger than the initial 0.5 even
for $\omega < \omega_c$. Now, a new comer with a random CP will have
a lower CP than the increased average of MCP. This in turn causes the
least fit site to be likely located next to the new comer's site,
resulting in avalanches of replacement activities. 

%%%%%%%%%%%%%%%%%%%%%%%%(Fig 3: OSDS)%%%%%%%%%%%%%%%%%%%%%%
\begin{figure}[t!] 
\includegraphics[width=8cm]{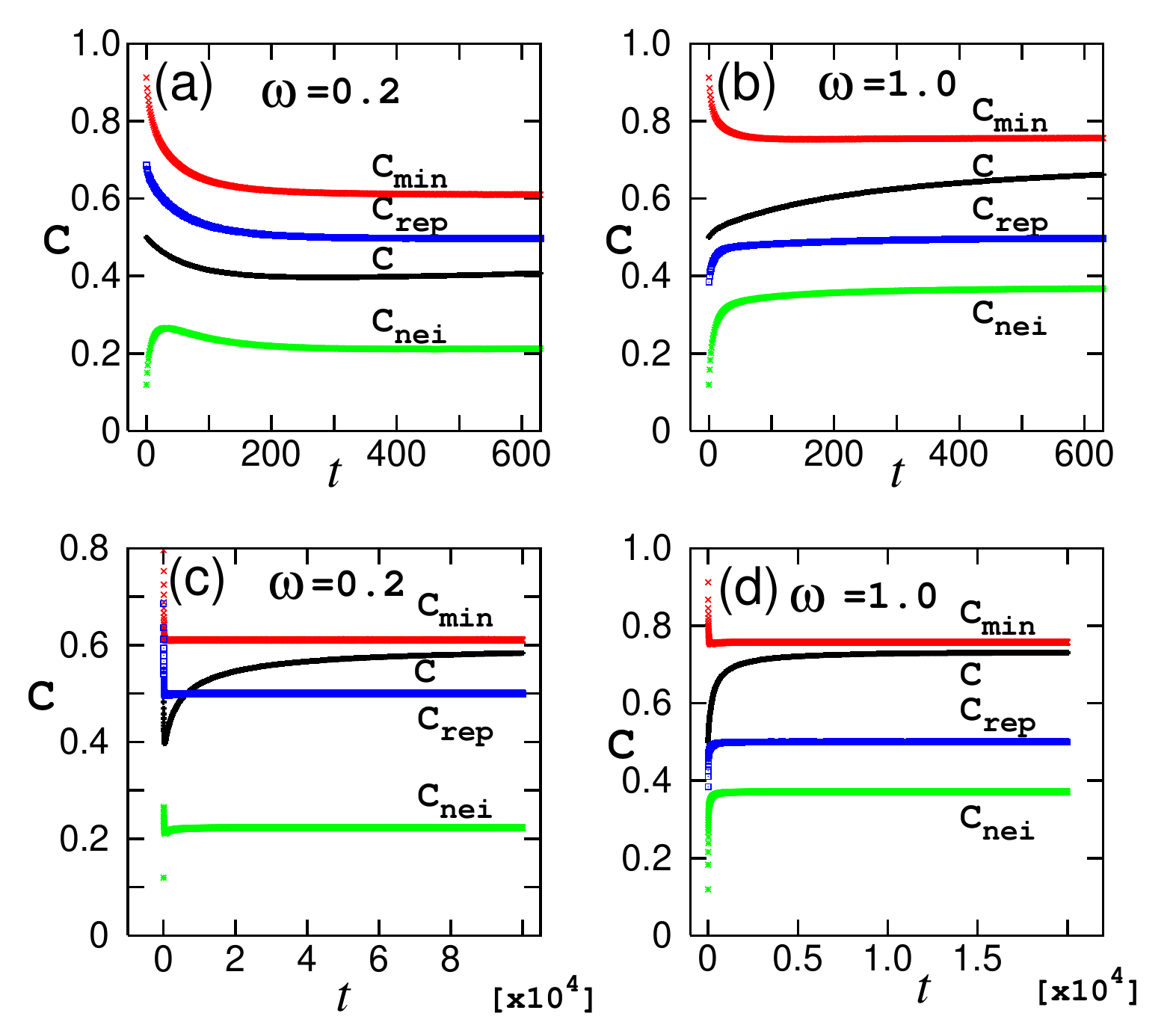} 
\caption{
  Evolution of the CPs of the least fit
  members, $C_{min}$, their neighbors, $C_{nei}$, 
  and members that are replaced, $C_{rep}$, are shown
  together with MCP, $C$, for (a,c) $\omega=0.2$,
  and (b,d) $\omega=1.0$.
  In (a) and (b), the initial transient behaviors are shown
  while overall behaviors are shown in (c) and (d). 
 The system size $N=128$ is used for all cases.
}
\label{f.Crep.t}
\end{figure}
%%%%%%%%%%%%%%%%%%%%%%%%%%%%%%%%%%%%%%%%%%%%%%%%%%%%%%%%%%%%

\section{Analysis of the initial dynamics}
We start from the population with random strategies. Hence, there is no
correlation between the CPs initially, and we may understand the initial
dynamics through the mean-field calculation. 
We first define the mean CP of the replacement sites (before
the replacement),
\beqa
C_{rep} \a= \frac{1}{1+2\omega} (C_{min} + 2\omega C_{nei}),
\eeqa
where mean-field dynamics can be easily analyzed.
Here, $C_{min}$ is the average of the CPs for the least fit members, 
and $C_{nei}$ is that for the neighbors of the least fit members. 
On average, CPs of $1+2\omega$ sites are updated each time.
Since the average of the newly assigned random cooperation rate
is 0.5, $C_{rep}$ satisfies, 
\beqa
 \frac{dC_{rep}}{dt} \a= \frac{1}{1+2\omega} \pas{(0.5 - C_{min}) + 2\omega ( 0.5 - C_{nei})} \nnn
                    \a= 0.5 - C_{rep}.
\label{e.Crep}
\eeqa
We measure $C_{rep}$ and present them in Fig.~\ref{f.Crep.t},
together with the CPs of the least fit members, $C_{min}$, 
that of the replaced members, $C_{rep}$, and
the MCP, $C$, for $\omega=0.2$ and $\omega=1.0$.
The $C_{rep}$ curves are, indeed, well described by Eq.~(\ref{e.Crep}).
If we represent the numerical solutions of Eq.~(\ref{e.Crep}) in the figure, 
they cannot be distinguished from the $C_{rep}$ curves from the
simulations because
they are almost identical. From Fig.~\ref{f.Crep.t}, we also see that 
$C_{rep}$ enters its steady value in a relatively short period of
time compared to $C$ and rapidly converges to its steady-state value of 0.5.
For $\omega=0.2$, the initial $C_{rep}$ is more than half
and hence decreases to the steady value of 0.5 
while it increases from the value below 0.5 for $\omega = 1.0$. 
For a sufficiently large system, 
the initial value of $C_{min}$ would be 1 while $C_{nei}$ is~0.
Hence, the initial value of $C_{rep}$ would be
$\frac{1}{1+2\omega}$, which is more than 0.5 
for $\omega < 1/2$. 
In this transient time of $C_{rep}$, 
the dynamics of MCP, $C$, would be mainly 
determined by the dynamics of $C_{rep}$.
Therefore, $C$ initially decreases for $\omega<1/2$ as does $C_{rep}$. 
However, after $C_{rep}$ reaches a steady value, 
the correlation of the replacement sites
mainly governs the dynamics, and $C$ begins to increase.
Let $m$ be the least fit member at time $t-1$;
then, at time $t$, $c_m$ is always updated, and 
$c_{j=m\pm 1}$ are updated with probability $\omega$.
After replacement, if the sum of the CPs at these three sites,
$s_{rep}(t) =  c_m(t) + \sum_{j=m\pm 1} c_j (t)$ (at the time $t$),
is small, at least one of $m-1$, $m$ or $m+1$ sites, 
is likely to have small fitness. 
Therefore, they will be easily replaced in a relatively short time. 
In other words, a new born member with small $s_{rep}(t)$
has a short lifetime and contributes less to the 
$C$ than those with large $s_{rep}(t)$. 
This mechanism makes $C$ increase 
up to (almost) $C_{min}$, and hence, the system becomes cooperative overall. 
Thus, according to our model, 
the emergence of cooperation is intrinsically related 
to the dynamics leading to  self-organized criticality (SOC).

\section{Self organized criticality}

%%%%%%%%%%%%%%%%%%%%%%%%(Fig 4: OSDS)%%%%%%%%%%%%%%%%%%%%%%
\begin{figure}[t!] 
\includegraphics[width=8.cm]{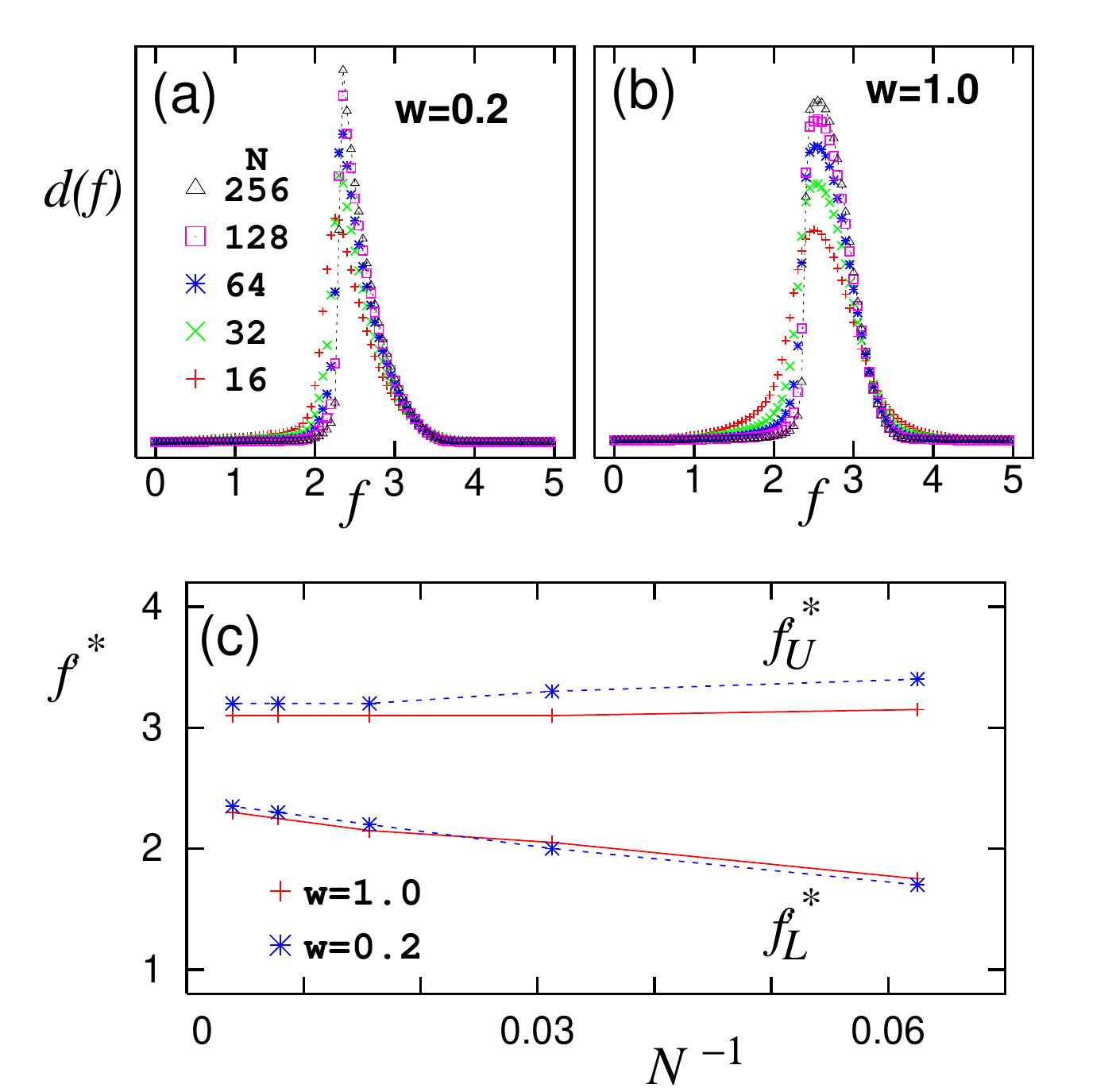} 
\caption[0]{
Fitness distribution $d(f)$ in the steady states for five
different system sizes of $N=16$, 32, 64, 128 and 256 with 
(a) $\omega=0.2$ and (b) $\omega=1.0$. 
The system size dependence of the effective lower and upper thresholds
$f_L^*$ and $f_U^*$ (defined in the text) are shown in (c).
Legends of (a) are also applied to (b).
} 
\label{f.4}
\end{figure}
%%%%%%%%%%%%%%%%%%%%%%%%%%%%%%%%%%%%%%%%%%%%%%%%%%%%%%%%%%%%

We now show that our model, in fact, drives 
the population into a SOC state as in the original BS model. 
We measure the distributions of avalanche sizes and 
distances between successive least fit sites in the steady states 
and show that they follow power-law distributions. 

Following Bak and Sneppen~\cite{Bak93}, we would like to define the
size of an avalanche as the number of subsequent replacements at 
the least fit sites below the lower threshold $f_L$ in its fitness value. 
The fitness distributions $d(f)$ share some
characteristics of the BS model~\cite{Bak93} although their overall shapes 
are quite different. A crucial similarity is that 
the fitness distribution $d(f)$ in the steady state
becomes zero for fitness $f$ smaller than a lower threshold 
$f_{L}$ as the system size goes to infinity. 

The fitness distributions in the steady states for five different system sizes 
are shown in Figs.~\ref{f.4}(a) and (b) for $\omega=0.2$ and
$\omega=1.0$. As the system sizes increase, the peak 
positions of the fitness distribution move to the right to high values,
and the peak widths become narrow. 
To estimate the threshold values $f_L$ and $f_U$, we define 
the effective lower [upper] threshold $f_L^*(N)$ [$f_U^*(N)$]  
as the $f$ value below [above] which the integrated distribution
is 5 percent. We plot them against $1/N$ in Fig.~\ref{f.4}(c)
for two different chain-death rates, $\omega=0.2$ and $\omega=1.0$. 
There are no noticeable differences in the thermodynamic values 
for the two $\omega$ values. Using linear fitting, we 
get rough estimates of the threshold values,
$f_L = 2.4 \pm 0.05$ and $f_U = 3.1 \pm 0.1 $, for
both $\omega$ values.

%%%%%%%%%%%%%%%%%%%%%%%%(Fig 5: OSDS)%%%%%%%%%%%%%%%%%%%%%%
\begin{figure}[t!] 
\includegraphics[width=8.cm]{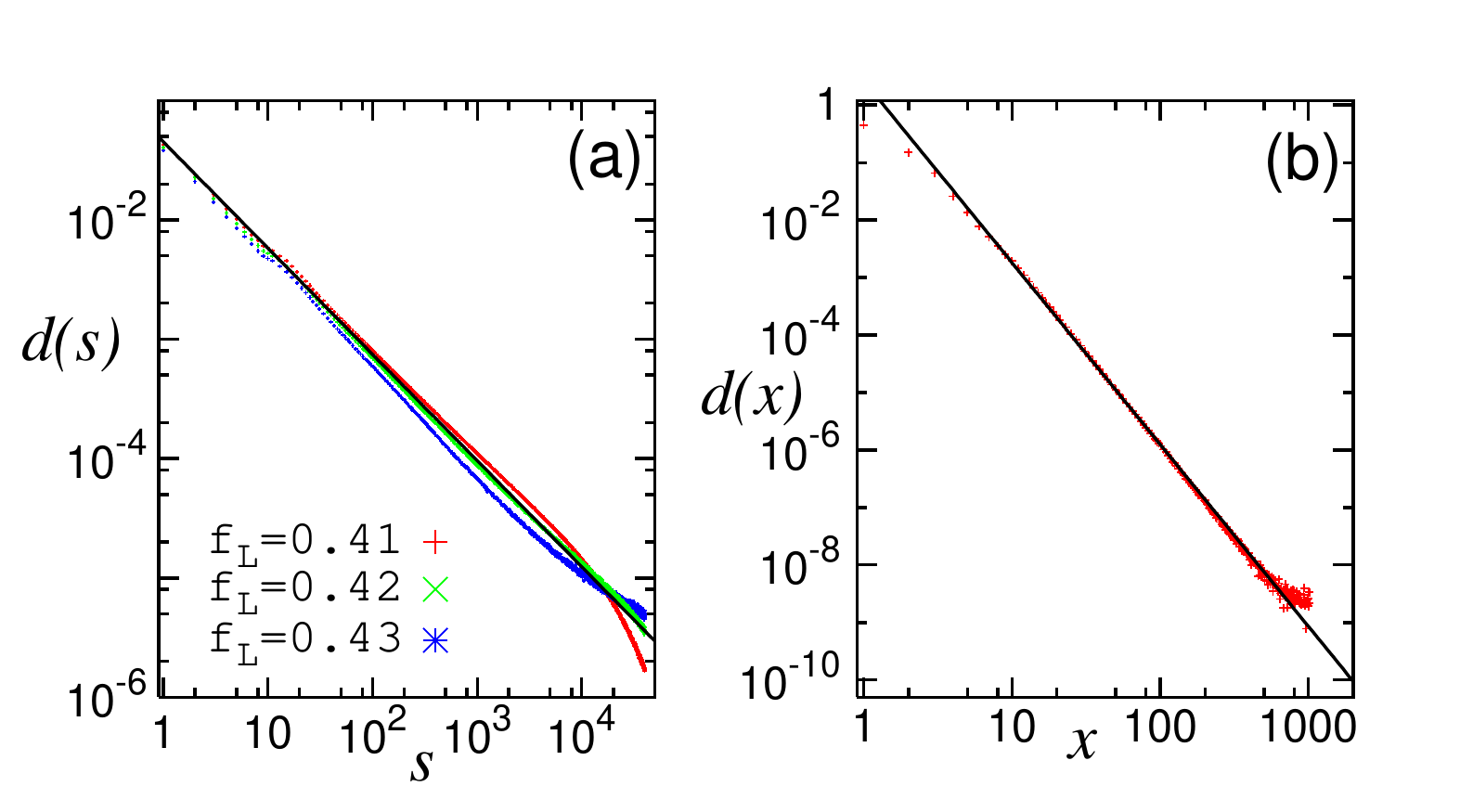} 
\caption[0]{
(a) Distributions $d(s)$ of avalanche sizes $s$.
Distributions with three different values of $f_L$, 
$f_L = 2.41$, $f_L = 2.42$, and $f_L = 2.43$ are 
measured in systems of $N=256$ in their steady states.
Data with $f_L = 2.42$ show a most persistent straight line
in the log-log scale fit, indicating the lower threshold 
$f_L = 2.42$ for the $N=256$ system with $\omega=1$. 
The black line is the least-squares fit of the data for $f_L = 2.42$
and is given in a form of $d(s) \sim s^{-\tau}$ with $\tau = 0.89 \pm 0.05$.  
(b) A distribution $d(x)$ of the distances between successive
minimum fitness sites in the steady states for the system of 
$N=2048$ with $\omega=1$. 
The black line is the least squares fit of the data 
in the form of $d(x) = a x^{-\alpha}$ with
$\alpha=3.17 \pm 0.03$.
}
\label{f.5}
\end{figure}
%%%%%%%%%%%%%%%%%%%%%%%%%%%%%%%%%%%%%%%%%%%%%%%%%%%%%%%%%%%%

For the avalanche size distribution $d(s)$, we need a more precise
value of $f_L$. We measure $d(s)$ with several 
different values of $f_L$ around the estimated value.
If the system is really in a SOC state, we expect the avalanche size
distribution $d(s)$ to show a power-law distribution,
for the exact value of $f_L$ for the given system. 
Figure~\ref{f.5}(a) shows the distribution of avalanche sizes
in a system of size $N=256$. 
We plot $d(s)$ against $s$ on a log-log scale with
three different values of $f_L$ around the value estimated 
from Fig.~\ref{f.4}(c) to pinpoint the threshold $f_L$. 
For $\omega=1.0$ shown in Fig.~\ref{f.5}(a), the avalanche size
distribution is well fit by a power-law with $f_L = 2.42$. 
It remains as a line in the log-log plot up to an avalanche size about 
20000, indicating power-law distributions $d(s) \sim s^{-\tau}$. 
The exponent obtained from a least-square fit 
of the form $d(s) = A y^{-\tau}$ is $\alpha=0.89 \pm 0.05$. 
This value is consistent with the known exponent of the 1D BS model~\cite{Bak93}.
The power law indicates that the evolution occurs in a
dynamical criticality~\cite{Bak93,Bak96B}. 
We measure the avalanche distributions for other $\omega$ and $b$
and found the critical exponent $\tau$ 
to be independent of the benefit-to-cost ratio $b$ or 
the chain-death probability $\omega$. 

We also measured the distance distribution between successive least fit sites.
Denoting the distance between successive minimum fitness sites by $y$,
we plot $d(y)$ in Fig.~\ref{f.5}(b). The distance distribution is
measured in the steady states for the system of $N=2048$ with $\omega=1$. 
When the distribution $d(y)$ is plotted against $y$ on a log-log
scale, it also becomes a line, indicating power-law distributions 
$d(y) \sim y^{-\alpha}$ with the slop $\alpha=3.17 \pm 0.03$.
This exponent is also consistent with the known exponent of the 1D BS model~\cite{Bak93}.
It is notable that our model belongs to the same universality class as 
the BS model in spite of the complexity in computing the fitness of 
members and the non-trivial dynamics of the population-fitness changes.

\section{Concluding Remarks}
We have considered the BS mechanism as a reproduction process with fitness
given by a PD game payoff on a network structure. Our observation may
have more natural implication in economical systems because
the BS process with chain bankruptcy is a more feasible scenario. 
It might be worthwhile 
analyzing weekly or monthly bankruptcy data and see if they follow a
power-law distribution as our study suggests. 

We have simulated our model with other values of the 
benefit-to-cost ratio $b$ and see that cooperation emerges in a
wide range of chain-death rates $\omega$, as long as $b$ is larger
than~1. In contrast to a common
belief, cooperation can emerge even with parameters that a population 
with random strategies decreases cooperation. This is possible because 
the BS mechanism
builds dynamical correlations that suppress the long-term survival of
non-cooperators even in the region where mean-field calculation
predicts a decrease in cooperators. The same dynamical correlation
leads to SOC in replacement activities with the same exponents as
the original BS model. 
The strategy space presented here is rather small. Mixed but
only history independent strategies are considered on 
a very simple population structure, a 1D lattice.
However, we speculate that our main results, the emergence 
of cooperation and SOC, are robust under variations in the 
population structure or the strategy space extension. 
In fact, the preliminary results with the extended strategy space show
that the emergence of cooperation appears more easily
and rapidly when the reactive strategies are included. 

\section{Acknowledgements}
This work was supported by the National Research Foundation of Korea
Grant funded by the Korean Government(MEST) (NRF-2010-0022474).
H.-C. J. would like to thank KIAS for the support during the visit. 

%\bibliography{All}
%\bibliographystyle{prsty}

\end{document}